\def\IEEETRAN{IEEEtran}

\def\DOUBLECOLUMN{
  \documentclass[twocolumn,twoside]{\IEEETRAN}
  \newcommand{\FIGWIDTH}{8.5cm}
  }
\def\SUBMITSTYLE{
  \documentclass[submit,11pt]{\IEEETRAN}
  \def\baselinestretch{1.66}
  \newcommand{\FIGWIDTH}{15cm}
  \newcommand{\appendices}{\appendix}
  }
\def\VANILLA{
  \documentclass[12pt,twoside]{article}
  \def\baselinestretch{1.25} 
  \usepackage{amsthm} 
  \pagestyle{myheadings}
  \setlength{\oddsidemargin}{0in}
  \setlength{\evensidemargin}{0in}
  \setlength{\textwidth}{6.5in}
  \setlength{\textheight}{8.5in}
  \setlength{\topmargin}{0in}
  \newcommand{\QED}{\qed}
  \newcommand{\FIGWIDTH}{10cm}
  \newcommand{\appendices}{\appendix}
  }

\DOUBLECOLUMN 

\usepackage{amsmath,amssymb,array,graphicx,color,epsfig,citesort}

\newtheorem{defn}{Definition}

\newtheorem{rmk}{Remark}

\newtheorem{thm}{Theorem}
\newtheorem{lem}{Lemma}

\newtheorem{cor}{Corollary}

\newcommand{\Real}{{\mathbb{R}}}
\newcommand{\Integer}{{\mathbb{Z}}}
\newcommand{\defas}{\overset{\mathrm{def}}{=}}
\newcommand{\set}[1]{{\mathcal{#1}}}

\DeclareMathOperator{\var}{var}

\newcommand{\deletethis}[1]{{}}
\newcommand{\comment}[1]{}

\newcommand{\rE}{{\mathsf{E}}}

\newcommand{\SNR}{{\mathsf{SNR}}}

\begin{document}
\markboth{Version \today}{Version \today}

\title{\emph{A Posteriori} Equivalence: A New Perspective for Design
  of Optimal Channel Shortening Equalizers}
\author{Raman Venkataramani~\IEEEmembership{Member,~IEEE,} and
  M.~Fatih Erden~\IEEEmembership{Member,~IEEE}\thanks{The authors are
    with Seagate Research, Pittsburgh, PA 15222. Email:
    ramanv@ieee.org, fatih.erden@seagate.com} }

\maketitle

\begin{abstract}
  The problem of channel shortening equalization for optimal detection
  in ISI channels is considered.  The problem is to choose a linear
  equalizer and a partial response target filter such that the
  combination produces the best detection performance.  Instead of
  using the traditional approach of MMSE equalization, we directly
  seek all equalizer and target pairs that yield optimal detection
  performance in terms of the sequence or symbol error rate. This
  leads to a new notion of \emph{a posteriori} equivalence between the
  equalized and target channels with a simple characterization in
  terms of their underlying probability distributions. Using this
  characterization we show the surprising existence an infinite family
  of equalizer and target pairs for which \emph{any} maximum \emph{a
    posteriori} (MAP) based detector designed for the target channel
  is simultaneously MAP optimal for the equalized channel. For
  channels whose input symbols have equal energy, such as $q$-PSK, the
  MMSE equalizer designed with a monic target constraint yields a
  solution belonging to this optimal family of designs.  Although,
  these designs produce IIR target filters, the ideas are extended to
  design good FIR targets.  For an arbitrary choice of target and
  equalizer, we derive an expression for the probability of sequence
  detection error. This expression is used to design optimal FIR
  targets and IIR equalizers and to quantify the FIR
  approximation penalty.
\end{abstract}

\begin{keywords}
  Intersymbol interference, linear equalization, channel shortening,
  partial response, target design, MAP detection, decision feedback.
\end{keywords}

\section{Introduction}
\label{sec:introduction}

The problem of designing channel shortening equalizers for
maximum-likelihood sequence detection in inter-symbol interference
(ISI) channels has been widely studied
\cite{messerschmitt.1974,falconer.magee.1973,lee.hill.1987,magee.1975,quereshi.newhall.1973}.
The function of the equalizer is to modify the channel response to
reduce the length of the ISI in the system thereby reducing the
complexity of the sequence detector.  Traditionally, the equalizer is
designed so that the equalized channel response approximates a
pre-specified short FIR sequence called the \emph{partial response}
(PR) target.
Two commonly studied classes of equalizers are the zero-forcing
equalizer (ZFE) and the minimum mean-squared error (MMSE) equalizer.
The ZFE forces the equalized channel response to match the target
response exactly. The undesired effect of zero forcing is that it
colors the noise spectrum and may amplify the noise significantly.  In
contrast, the MMSE equalizer minimizes the variance of the
equalization error, but the error is signal dependent.  In both cases
the goal is to make the equalized channel response \emph{close} to the
target response. However, the ultimate goal of the channel shortening
equalization ought to be a detection performance measure such as the
sequence or symbol error rate.

In this work we take revisit the problem equalizer design in the
context of optimal (MAP) detection of the input.  
The main contribution of this paper is a new perspective for the
problem of channel shortening equalization in terms of the underlying
\emph{a posteriori} probabilities (APPs) rather than the traditional
approach of using the MMSE equalization error as the criterion
\cite{DBLP:journals/tit/Al-DhahirC96a,messerschmitt.1974,falconer.magee.1973,lee.hill.1987,magee.1975,quereshi.newhall.1973}.
We pose the question: \emph{In what sense should the target channel be
  equivalent to the equalized channel to achieve best detection
  performance?} The answer to this question naturally leads us to a
new notion of \emph{a posteriori equivalence} (APE) between the
equalized channel and the target channel. We show that this form of
equivalence, which is expressed in terms of their underlying \emph{a
  posteriori} probabilities, guarantees no performance loss due to
equalization compared to the optimal detector for the original
channel.  This result thus provides a new recipe for equalizer and
target design, which is different from the heuristic approach of
matching the responses of the target and the equalized channel. We
also prove that there is a family of IIR equalizers and targets which
guarantee APE.

This paper is organized as follows.  In
Sections~\ref{sec:sequence.detection} and
\ref{sec:linear.equalization} we review the background material on
optimal sequence detection and linear equalization. In
Section~\ref{sec:sequence.detection.equalized} we consider the problem
of sequence detection for the equalized channel.  We present our main
theoretical results including \emph{a posteriori} equivalence and its
algebraic characterization. In
Section~\ref{sec:practical.considerations} we consider practical
implications of our results. In particular, we show that the MMSE
equalizer designed with a monic target constraint yields an optimal
solution for ISI channels when the input symbols have equal energy.
Unfortunately, the equivalence conditions usually hold only for IIR
targets, making the results somewhat useless for channel shortening.
However, in Section~\ref{sec:FIR.design} we extend the results to FIR
target design where we seek the best FIR target and IIR equalizer with
a small but acceptable ``FIR approximation penalty.'' We derive an
expression for the sequence detection error rate, and use this as a
performance measure for the filter design. For simplicity of the
analysis we consider only IIR equalizers. The problem of FIR equalizer
design would entail the additional task of optimizing the processing
delay.  We refer the reader to
\cite{DBLP:journals/tit/Al-DhahirC96a,dhahir.cioffi.1995,dhahir.2000}
for the problem on optimizing the processing delay for systems using
MMSE equalization.  A similar analysis related to FIR equalizer design
would be equally important in our problem, but is beyond the scope of
this paper.  Finally, in Section~\ref{sec:examples} we apply our
theory to an example ISI channel with binary and non-binary inputs to
confirm our predictions through computer simulation.

\subsection{Definitions and Notation}

Let $\boldsymbol{a}$ denote a discrete-time sequence $\{a_n:
n\in\Integer \}$. If $\boldsymbol{a}$ has finite energy its
discrete-time Fourier transform is defined as
\[
\set{F} \{\boldsymbol{a}\}=A(\omega)= \sum_n a_n e^{-j n
\omega}.
\]
The convolution of two sequences $\boldsymbol{a}$ and $\boldsymbol{b}$
is denoted by $\boldsymbol{c}=\boldsymbol{a}\star
\boldsymbol{b}$:
\[
c_n =\sum_m a_mb_{n-m}.
\]
Let $\boldsymbol{\delta}$ denote the discrete delta function:
$\delta_n=0$ for $n\neq 0$ and $\delta_0=1$. 
Define the inner product between two sequences $\boldsymbol{a}$ and
$\boldsymbol{b}$ as
\[
\langle \boldsymbol{a}, \boldsymbol{b} \rangle = \sum_n a_n^* b_n = 
\frac{1}{2\pi} \int_{-\pi}^{\pi} A^*(\omega)B(\omega)d\omega
\]
where $*$ denotes complex conjugation for scalars or
conjugate-transposition for matrices. Thus, the norm of $\boldsymbol{a}$
is
\[
\|\boldsymbol{a}\| = \langle \boldsymbol{a}, \boldsymbol{a} \rangle ^{1/2}.
\]

Given a sequence $\boldsymbol{a}$, let $\ddot{\boldsymbol{a}}$ be
obtained by time-reversal and conjugation of $\boldsymbol{a}$, i.e.,
\[
\ddot{a}_n = a_{-n}^*.
\]
The Fourier transform of $\ddot{\boldsymbol{a}}$ is
$A^*(\omega)$. Thus, we readily obtain the following identity:
\begin{align}
\langle \boldsymbol{a}\star \boldsymbol{b}, \boldsymbol{c} \rangle =
\langle \boldsymbol{b}, \ddot{\boldsymbol{a}} \star \boldsymbol{c}
\rangle \label{eq:adjoint}
\end{align}
i.e., the adjoint of the convolution operation with $\boldsymbol{a}$
is convolution with $\ddot{\boldsymbol{a}}$.

Let $\boldsymbol{x}$ and $\boldsymbol{y}$ denote real or complex
stationary random processes. The cross-correlation function is defined
by
\[
r^{xy}_n = \zeta^{-1} \rE (x_{m+n} y^*_{m})
\]
where $\rE(\cdot)$ denotes expectation and $\zeta$ is the number of
real dimensions per sample, i.e., $\zeta=1$ for real processes and
$\zeta=2$ for complex ones.  The autocorrelation of $\boldsymbol{x}$ is
obtained by setting $\boldsymbol{y}=\boldsymbol{x}$.  The power
spectral density of $\boldsymbol{x}$ is $S_x(\omega) = \set{F}
\{\boldsymbol{r}^{xx}\}$.  We write $\boldsymbol{x} \perp
\boldsymbol{y}$ if $\boldsymbol{r}^{xy}=0$. 

\subsection{ISI Channel Model}

Consider the following discrete-time model for a real or
complex-valued linear time invariant system
\begin{align}
\boldsymbol{y}= \boldsymbol{h}\star \boldsymbol{x} + \boldsymbol{w}
\label{eq:channel.model}
\end{align}
where $\boldsymbol{x}=\{x_m\}$ is the input to the channel,
$\boldsymbol{h}=\{h_m\}$ is the channel impulse response and
$\boldsymbol{w}=\{w_n\}$ is additive white Gaussian noise with
$S_w(\omega)=\sigma_w^2$. Assume that $\boldsymbol{h}$ has finite
energy but is possibly non-causal and infinite.  The channel model
(\ref{eq:channel.model}) is usually the base-band representation after
whitened matched filtering \cite{book:proakis} and describes a variety
of communication systems.

In the case of complex channels, the noise is assumed to be
circularly symmetric. Thus, the real and imaginary components of the
noise samples are independent with variance $\sigma_w^2$. Let the
input power spectral density be $S_x(\omega)$. As a special case we
also shall consider independent and identically distributed (IID)
inputs with $S_x(\omega)=1$.  An example for the input symbol set is
the $Q$-phase PSK constellation,
\[
\set{C} =\{\sqrt{2} e^{j2\pi q/Q}: q=0,\dots, Q-1\}
\]
in the complex case or the BPSK (bipolar binary) constellation
$\set{C} = \{-1, +1\}$ in the real case.

\section{Optimal Sequence Detection}
\label{sec:sequence.detection}

Suppose that a message $\boldsymbol{x}=\{x_m: m=0,\dots, M-1\}$ of
finite length $M$ symbols is transmitted through the channel
(\ref{eq:channel.model}). The received signal is given by
\begin{align}
y_n=\sum_{m=0}^{M-1} h_{n-m} x_{m} +w_n.
\label{eq:channel.model.2}
\end{align}
Since the additive noise is white Gaussian, we have
\begin{align}
P(\boldsymbol{y}|\boldsymbol{x}) \propto \exp 
\Big( -\frac{D(\boldsymbol{y}, \boldsymbol{x})}
{2\sigma_w^2}\Big)
\label{eq:p(y|x)}
\end{align}
where
\begin{align}
D(\boldsymbol{y}, \boldsymbol{x}) = \sum_n \Big|y_n - 
\sum_{m=0}^{M-1} h_{n-m} x_{m}\Big|^2
\label{eq:distance.cost}
\end{align}
with the summation over $n$ carried over the finite region of
interest where the samples $y_n$ are available.

Given the output sequence $\boldsymbol{y}$, the maximum \emph{a
posteriori} (MAP) estimate of $\boldsymbol{x}$ is given by
\begin{align}
\hat{\boldsymbol{x}} &\defas \arg\max_{\boldsymbol{x}} 
P(\boldsymbol{x}|\boldsymbol{y}) = \arg\max_{\boldsymbol{x}}
P(\boldsymbol{y}|\boldsymbol{x}) P(\boldsymbol{x}) \nonumber \\ 
&=\arg\min_{\boldsymbol{x}} \Big(\frac{D(\boldsymbol{y},
\boldsymbol{x})}{2\sigma_w^2} - \log P(\boldsymbol{x}) \Big)
\label{eq:MAP.optimal.rule}
\end{align}
where $P(\boldsymbol{x})$ is the prior probability distribution on
$\boldsymbol{x}$.  If this distribution is uniform, then
(\ref{eq:MAP.optimal.rule}) reduces to maximum-likelihood (ML)
detection of the input sequence:
\begin{align}
\hat{\boldsymbol{x}} = \arg\min_{\boldsymbol{x}} D(\boldsymbol{y}, 
\boldsymbol{x}) = \|\boldsymbol{y} - \boldsymbol{h}\star \boldsymbol{x}\|^2.
\label{eq:ML.optimal.rule}
\end{align}

Unfortunately, the direct use of the above expression is limited due
to its computational complexity which grows exponentially with the
length of the ISI.  However, when $h_n$ is a short FIR sequence, the
above cost function can be minimized exactly and
computationally efficiently using the Viterbi algorithm
which was originally devised to decode convolutional codes 
\cite{viterbi.1967,viterbi.1971,book:proakis}.

\section{Review of Linear Equalization}
\label{sec:linear.equalization}

In order to implement the Viterbi algorithm to solve the ML sequence
detection (\ref{eq:ML.optimal.rule}) with manageable complexity, we
need to reduce the length of the ISI in the system. This is usually
accomplished by using a linear equalizer to condition the channel
response to match a pre-specified \emph{target response}. When the
target is a short FIR filter, it is called a \emph{partial response}
(PR) target. The Viterbi detector operates on the equalized channel to
perform sequence detection pretending that the samples were the output
of a hypothetical target channel.

\begin{figure}[htb]
\begin{minipage}[b]{1.0\linewidth}
  \centering
  \centerline{{\epsfig{figure=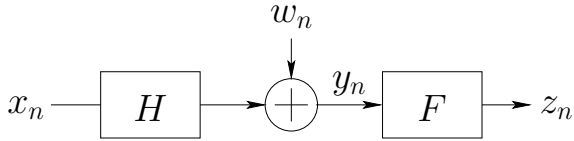,width=\FIGWIDTH}}}
  \caption{The equalized channel}
  \label{fig:equalized.system}
\end{minipage}
\end{figure}

Let $\boldsymbol{f}=\{f_n\}$ and $\boldsymbol{g}=\{g_n\}$ denote the
equalizer and target filters respectively. For the moment assume that the
target is fixed.  Fig.~\ref{fig:equalized.system} illustrates the
system with an equalizer whose output is
\begin{align}
\boldsymbol{z} &=\boldsymbol{f}\star \boldsymbol{y} = 
\boldsymbol{f}\star \boldsymbol{h}\star \boldsymbol{x} + 
\boldsymbol{f}\star \boldsymbol{w} \nonumber \\
&= \boldsymbol{l}\star \boldsymbol{x} + \boldsymbol{u}
\label{eq:equalized.channel}
\end{align}
where $\boldsymbol{l}=\boldsymbol{f} \star \boldsymbol{h}$ is the the
response of the equalized channel and
$\boldsymbol{u}=\boldsymbol{f}\star \boldsymbol{w}$ is the output
noise whose power spectral density is $S_u(\omega) = |F(\omega)|^2
S_w(\omega) =\sigma_w^2 |F(\omega)|^2$.

\begin{defn}
  The \emph{target channel} is a hypothetical channel defined by
\begin{align}
\tilde{\boldsymbol{z}}=\boldsymbol{g} \star \boldsymbol{x} + 
{\boldsymbol{v}}
\label{eq:target.channel}
\end{align}
where $\boldsymbol{x}$ is the input, $\boldsymbol{v}$ is additive
white Gaussian noise with $S_{v}(\omega)=\sigma_v^2$, and
$\tilde{\boldsymbol{z}}$ is the output.
\end{defn}

\begin{figure}[htb]
\begin{minipage}[b]{1.0\linewidth}
  \centering
  \centerline{{\epsfig{figure=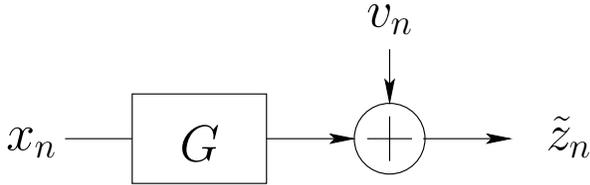,width=\FIGWIDTH}}}
  \caption{The target channel}
  \label{fig:target.channel}
\end{minipage}
\end{figure}

The original channel with the equalizer is illustrated in
Fig.~\ref{fig:equalized.system} and the target channel that
approximates it is shown in Fig.~\ref{fig:target.channel}.
Traditionally, the equalizer and target are designed to make the
equalized channel response $\boldsymbol{l}$ \emph{close} to target
$\boldsymbol{g}$, while keeping the noise white.


\subsection{Zero Forcing Equalizer (ZFE)}
The ZFE modifies the channel response to match the
target filter exactly, i.e., $\boldsymbol{l}=\boldsymbol{g}$. Thus,
in the frequency domain, the equalizer is given by
\begin{align}
F(\omega) = \frac{G(\omega)}{H(\omega)}.
\end{align}

The spectral density of the noise $\boldsymbol{u}$ is 
\begin{align}
S_u(\omega) = |F(\omega)|^2 S_w(\omega) =
\frac{|G(\omega)|^2}{|H(\omega)|^2}\sigma_w^2.
\label{eq:ZFE.noise.PSD}
\end{align}
An undesirable problem with zero-forcing equalization is that when the
channel response $|H(\omega)|$ has a spectral null or attains very
small values, the equalized noise is highly colored and has large
variance. The ZFE is rarely used for this reason.

\subsection{Minimum Mean Squared Error (MMSE) Equalizer}
A widely used equalizer in practical systems is the MMSE equalizer
which is designed to minimize the variance of the equalization error
$\boldsymbol{e}$ defined as
\begin{align}
\boldsymbol{e} \defas 
\boldsymbol{g}\star \boldsymbol{x} -\boldsymbol{f}\star \boldsymbol{y}.
\label{eq:equalization.error}
\end{align}
The MMSE equalizer ensures that $\boldsymbol{e} \perp
\boldsymbol{y}$, which yields
\begin{align}
F(\omega)=\frac{S_x(\omega)H^*(\omega)G(\omega)}{|H(\omega)|^2S_x(\omega) +
\sigma_w^2}
\label{eq:MMSE.equalizer}
\end{align}
where $S_x(\omega)$ is the power spectral densities of the input
$\boldsymbol{x}$.
The spectral density of the estimation error is given by
\begin{align}
S_e(\omega) = \frac{|G(\omega)|^2 S_x(\omega) \sigma_w^2}
{|H(\omega)|^2S_x(\omega) +\sigma_w^2}.
\label{eq:MMSE.noise.PSD}
\end{align}
The advantage of the MMSE design over the ZFE is that the spectrum of
the MMSE noise (\ref{eq:MMSE.noise.PSD}) is less colored and always
smaller than the ZFE noise (\ref{eq:ZFE.noise.PSD}) and spectral nulls
in $H(\omega)$ cause no problems.  However, $\boldsymbol{e}$ is signal
dependent, which may cause the Viterbi detection to be suboptimal.

\subsection{Target Design}
Instead of choosing a fixed target, we seek the best target of a fixed
length. In practice, the target is usually designed for an MMSE
equalizer. Thus, we minimize the variance of the MMSE equalization
error (\ref{eq:MMSE.noise.PSD}):
\begin{align}
\min_{\boldsymbol{g}} \frac{1}{2\pi}\int_{-\pi}^{\pi} S_e(\omega)d\omega 
\label{eq:cost}
\end{align}
where the target $\boldsymbol{g}$ is assumed to have length $L$:
\[
\boldsymbol{g} =\{g_0,g_1,\dots, g_{L-1} \}.
\]
The resulting cost function is a simple quadratic function of the
target filter taps.  Clearly, with no further constraints on
$\boldsymbol{g}$ we obtain the trivial solution
$\boldsymbol{g}=0$. Therefore, an additional constraint is imposed on
$\boldsymbol{g}$ such as the \emph{unit-energy constraint}
\begin{align}
\sum_n g_n^2 = 1
\label{eq:unit.energy.constraint}
\end{align}
or the \emph{monic constraint}
\begin{align}
g_0 = 1.
\label{eq:monic.constraint}
\end{align}
or sometimes the \emph{unit-tap constraint} $g_k=1$ for some $k$.  In
each of these cases, the optimal target, known as the
\emph{generalized partial response} (GPR) target, is found easily by
solving (\ref{eq:cost}) subject to the appropriate constraints.

For illustrative purposes, we derive the solution to the monic design
in the IIR limit ($L\to\infty$), where the problem can be expressed in
the frequency domain as
\begin{align}
\min_{\boldsymbol{g}} = \frac{1}{2\pi} 
\int_{-\pi}^{\pi}\frac{|G(\omega)|^2S_x(\omega)\sigma_w^2}
{|H(\omega)|^2S_x(\omega) + \sigma_w^2}d\omega
\label{eq:MMSE.cost.function}
\end{align}
over all causal targets $\boldsymbol{g}$ with $g_0=1$. 

The causal and monic constraint on $\boldsymbol{g}$ is cumbersome to
express directly in the frequency domain. However, we know that among
all the causal and stable spectral factors of $Q(\omega) =
|G(\omega)|^2$ the value of $g_0$ is maximized for the minimum-phase
factor \cite{oppenheim:schafer:1975}.  This maximum value is given by
\[
\log g_0 = \frac{1}{2\pi} \int_{-\pi}^{\pi} \log Q(\omega) d\omega.
\]
Therefore, we rewrite the optimization (\ref{eq:MMSE.cost.function})
in terms of $Q(\omega)$ as
\[
\min_{\boldsymbol{g}} \frac{1}{2\pi} 
\int_{-\pi}^{\pi}\frac{Q(\omega)S_x(\omega)\sigma_w^2}
{|H(\omega)|^2S_x(\omega) + \sigma_w^2}d\omega
\]
such that
\begin{align}
\frac{1}{2\pi} \int_{-\pi}^{\pi} \log Q(\omega) d\omega = 0.
\label{eq:CV.constraint}
\end{align}
The Lagrangian
\[
\set{L}(\boldsymbol{q}, \lambda)= \int_{-\pi}^{\pi}\frac{Q(\omega)
S_x(\omega)\sigma_w^2d\omega}{|H(\omega)|^2S_x(\omega) + \sigma_w^2} - \lambda 
\int_{-\pi}^{\pi} \log Q(\omega) d\omega
\]
is stationary at the solution.  Using calculus of variations, we obtain
\begin{align}
|G(\omega)|^2 &= Q(\omega) = \lambda\frac{|H(\omega)|^2S_x(\omega) +
\sigma_w^2} {S_x(\omega)\sigma_w^2} \nonumber \\ &=
\frac{\lambda}{\sigma_w^2} |H(\omega)|^2 + \frac{\lambda}{S_x(\omega)}
\label{eq:target.for.monic.general}
\end{align}
where the Lagrange multiplier $\lambda$ is chosen to satisfy
(\ref{eq:CV.constraint}):
\[
\lambda = {\exp \Big( -\frac{1}{2\pi}
\int_{-\pi}^{\pi} \log\Big( \frac{|H(\omega)|^2S_x(\omega) + \sigma_w^2}
{S_x(\omega)\sigma_w^2}\Big)d\omega\Big)}.
\]
The optimal $G(\omega)$ is the causal minimum-phase spectral factor of
$Q(\omega)$, and the MMSE equalizer (\ref{eq:MMSE.equalizer}) reduces
to
\begin{align}
F(\omega) = \frac{\lambda}{\sigma_w^2} \frac{H^*(\omega)}{G^*(\omega)}
\label{eq:equalizer.for.monic.general}
\end{align}
The spectrum of the estimation error (\ref{eq:MMSE.noise.PSD}) is
white for this solution: 
\begin{align}
S_e(\omega)=\lambda.
\label{eq:white.error}
\end{align}
Henceforth, we refer to this solution as the \emph{monic design} or
\emph{monic solution} implicitly associating the optimal target with
the MMSE equalizer.  For the special case of zero-mean IID inputs with
$S_x(\omega)=1$, the above solution reduces to
\begin{align}
F(\omega) &=\frac{H^*(\omega)G(\omega)}{|H(\omega)|^2+\sigma_w^2}
\label{eq:MMSE.equalizer.for.monic.IID} \\
|G(\omega)|^2 &= \frac{\lambda}{\sigma_w^2}(|H(\omega)|^2 + \sigma_w^2)
\label{eq:target.for.monic.IID}
\end{align}
Coincidentally, this solution is related to the linear MMSE decision
feedback equalizer (DFE) for the given ISI channel
\cite{salz.1973,falconer.foschini.1973,belfiore.park.1979,cioffi.etal.DFE1,cioffi.etal.DFE2}.
The MMSE-DFE structure is optimal in achieving the capacity for an ISI
channel with additive white Gaussian noise
\cite{cioffi.etal.DFE1,cioffi.etal.DFE2,shamai.laroia.1996,guess.varanasi.2005}.
However, it is not immediately obvious or even always true that the
above equalizer and target filters would be optimal for sequence
detection of (non-Gaussian) input symbols.

As a caveat we reiterate that the sequence detection is not meant to
be implemented with decision feedback. We still use the Viterbi
algorithm or a MAP based algorithm such as the forward-backward
algorithm to compute the symbol \emph{a posteriori} probabilities
(APPs).  It has been observed that the monic design performs better in
detection than other design criteria such as the energy constraint
(\ref{eq:unit.energy.constraint}) or the \emph{unit-tap constraint} on
the target.  In the following section, we shall formally prove this
conjecture.

In practice, we need to design FIR equalizers and targets with unknown
channel and noise characteristics. In this case the second order
statistics of the channel input and output are estimated using
training and subsequently used to design FIR filters.
The solutions to these problems for the various target constraints is
described in
\cite{messerschmitt.1974,DBLP:journals/tit/Al-DhahirC96a,moon.zeng}.
We point out that this method is also applicable if the noise is
colored 
because the design ensures that the noise whitening is automatically
absorbed into the equalizer $\boldsymbol{f}$.  

\section{Sequence Detection for the Equalized Channel}
\label{sec:sequence.detection.equalized}

Traditionally, the sequence detection is performed in two steps.  The
first step is to equalize the channel output. The next step is to
perform the detection pretending that the equalizer output
$\boldsymbol{z}$ (Fig.~\ref{fig:equalized.system}) were the output of
the hypothetical target channel (Fig.~\ref{fig:target.channel}).  In
other words, although the sequence detector is optimally designed for
the target channel it is, in reality, applied to the equalized
channel. In this section we consider the performance of such a
detector. For simplicity of analysis we assume that the target and
equalizer are IIR and the target is causal.  We consider the design of
FIR targets in Section \ref{sec:FIR.design}.

Consider the system described by (\ref{eq:equalized.channel}),
restated below:
\[
\boldsymbol{z} = \boldsymbol{l}\star \boldsymbol{x} + \boldsymbol{f}\star 
\boldsymbol{w}.
\]
By design, the above channel approximates the target channel
(\ref{eq:target.channel}).  The conditional probability of the output
of the target channel is
\begin{align}
P(\tilde{\boldsymbol{z}}|\boldsymbol{x}) \propto \exp 
\Big( -\frac{\tilde{D}(\boldsymbol{z}, \boldsymbol{x})}
{2\sigma_v^2}\Big)
\label{eq:p(z.tilde|x)}
\end{align}
where
\begin{align}
\tilde{D}(\tilde{\boldsymbol{z}}, \boldsymbol{x}) = \sum_n 
\Big| \tilde{z}_n - \sum_{m=0}^{M-1} g_{n-m} x_{m} \Big|^2.
\label{eq:alternate.distance}
\end{align}
with the summation over $n$ carried over a finite region of interest
where the samples of $\tilde{\boldsymbol{z}}$ are available. The
following result provides an alternate expression for
$\tilde{D}(\tilde{\boldsymbol{z}}, \boldsymbol{x})$ which will be
useful in proving a form of equivalence between the target
channel and the equalized channel.
\begin{lem}
Suppose the equalizer $\boldsymbol{f}$ and target $\boldsymbol{g}$ are
chosen such that $\ddot{\boldsymbol{g}}\star \boldsymbol{f} = \alpha
\ddot{\boldsymbol{h}}$ for some $\alpha>0$, then
\begin{align*}
\tilde{D}(\boldsymbol{z}, \boldsymbol{x}) - \|\boldsymbol{z}\|^2
 &= \langle \boldsymbol{x}, \boldsymbol{s}\star \boldsymbol{x}
 \rangle+\alpha (D(\boldsymbol{y}, \boldsymbol{x}) - \|\boldsymbol{y}\|)^2
\end{align*}
where $\boldsymbol{s} = \ddot{\boldsymbol{g}}\star \boldsymbol{g}-
\alpha \ddot{\boldsymbol{h}}\star \boldsymbol{h}$.
\label{lem:part.result}
\end{lem}
\begin{proof}
We begin by expanding $\tilde{D}(\boldsymbol{z}, \boldsymbol{x})$ as follows
\begin{align*}
\tilde{D}(\boldsymbol{z}, \boldsymbol{x}) &= 
\|\boldsymbol{z}-\boldsymbol{g}\star \boldsymbol{x}\|^2 \\
&= \|\boldsymbol{z}\|^2 -2\Re \langle 
\boldsymbol{g}\star \boldsymbol{x}, \boldsymbol{z} \rangle+ 
\langle \boldsymbol{g}\star
\boldsymbol{x}, \boldsymbol{g}\star \boldsymbol{x}\rangle \\
&= \|\boldsymbol{z}\|^2 -2\Re \langle \boldsymbol{x} , 
\ddot{\boldsymbol{g}}\star
\boldsymbol{f} \star \boldsymbol{y}
\rangle + \langle 
\boldsymbol{x}, \ddot{\boldsymbol{g}}
\star\boldsymbol{g}\star \boldsymbol{x}\rangle 
\end{align*}
where $\Re$ denotes the real part. The last step follows by applying
(\ref{eq:adjoint}) and using $\boldsymbol{z}=\boldsymbol{f}\star
\boldsymbol{y}$. Using the hypothesis that $\ddot{\boldsymbol{g}}\star
\boldsymbol{f} = \alpha
\ddot{\boldsymbol{h}}$ we obtain
\begin{align}
\tilde{D}(\boldsymbol{z}, \boldsymbol{x})
&= \|\boldsymbol{z}\|^2 -2\alpha \Re \langle 
\ddot{\boldsymbol{h}} \star \boldsymbol{y},
\boldsymbol{x} \rangle + \langle 
\boldsymbol{x}, \ddot{\boldsymbol{g}}
\star\boldsymbol{g}\star \boldsymbol{x}\rangle .
\label{eq:part1}
\end{align}
Meanwhile, a similar argument shows that
\begin{align}
D(\boldsymbol{y},\boldsymbol{x}) &=
\|\boldsymbol{y}-\boldsymbol{h}\star \boldsymbol{x} \|^2 \nonumber\\
&=\|\boldsymbol{y}\|^2 -2\Re \langle 
\ddot{\boldsymbol{h}} \star \boldsymbol{y},
\boldsymbol{x} \rangle + \langle 
\ddot{\boldsymbol{h}}
\star\boldsymbol{h}\star \boldsymbol{x}, \boldsymbol{x}
\rangle.
\label{eq:part2}
\end{align}
From (\ref{eq:part1}) and (\ref{eq:part2}), we obtain the desired
result
\begin{align*}
\tilde{D}(\boldsymbol{z}, \boldsymbol{x}) - \|\boldsymbol{z}\|^2
 &= \langle \boldsymbol{x}, \boldsymbol{s}\star \boldsymbol{x}
 \rangle+\alpha (D(\boldsymbol{y}, \boldsymbol{x}) - \|\boldsymbol{y}\|)^2
\end{align*}
where $\boldsymbol{s} = \ddot{\boldsymbol{g}}\star \boldsymbol{g}-
\alpha \ddot{\boldsymbol{h}}\star \boldsymbol{h}$.
\end{proof}

\subsection{Equivalence of Equalized Channel and Target Channel}

We now interpret Lemma~\ref{lem:part.result} in terms of the
underlying probability distributions.  Let upper-case letters denote
random variables and lower-case letters denote realizations of these
random variables.  Suppose that $F(\omega)$ is a stable filter, i.e.,
it has no spectral nulls or singularities.  Then,
$\boldsymbol{z}=\boldsymbol{f}\star\boldsymbol{y}$ is
invertible. Hence, for the equalized channel
\begin{align*}
P(\boldsymbol{x}|\boldsymbol{z}) &= P(\boldsymbol{x}|\boldsymbol{y}) \propto P(\boldsymbol{x}) P(\boldsymbol{y}|\boldsymbol{x}) \\
&\propto P(\boldsymbol{x}) \exp\Big(-\frac{D(\boldsymbol{y},\boldsymbol{x})}
{2\sigma_w^2}\Big).
\end{align*}
where the constants of proportionality above (and henceforth) are always
independent of $\boldsymbol{x}$. Using Lemma~\ref{lem:part.result} and
noting that $\boldsymbol{y}$ and $\boldsymbol{z}$ are constants, we
obtain
\begin{align}
P(\boldsymbol{x}|\boldsymbol{z}) &\propto P(\boldsymbol{x}) 
\exp\Big( -\frac{\tilde{D}(\boldsymbol{z},\boldsymbol{x})}{2\alpha\sigma_w^2}
+\frac{\langle \boldsymbol{x}, \boldsymbol{s}\star \boldsymbol{x}
 \rangle}{2\alpha \sigma_w^2}\Big).
\label{eq:pdf1}
\end{align}

Suppose that the hypothetical target channel is assigned an input
prior distribution $\tilde{P}(\boldsymbol{x})$ which is possibly
different from $P(\boldsymbol{x})$.  The \emph{a posteriori}
probability of $\boldsymbol{x}$ is
\begin{align}
P(\boldsymbol{x}|\tilde{\boldsymbol{z}}) &\propto 
\tilde{P}(\boldsymbol{x}) P(\tilde{\boldsymbol{z}}|\boldsymbol{x}) 
\propto \tilde{P}(\boldsymbol{x}) \exp 
\Big(-\frac{\tilde{D}(\tilde{\boldsymbol{z}},
\boldsymbol{x})}{2\sigma_v^2}\Big). 
\label{eq:MAP.optimal.rule.2}
\end{align}
Comparing (\ref{eq:pdf1}) and (\ref{eq:MAP.optimal.rule.2}), we see
that by setting the noise variance $\sigma_v^2$ and the input prior
distribution $\tilde{P}(\boldsymbol{x})$ of the target channel
(\ref{eq:target.channel}) to
\begin{align}
\sigma_v^2 &\defas \alpha\sigma_w^2 \label{eq:sigma_v^2} \\
\tilde{P}(\boldsymbol{x}) &\propto P(\boldsymbol{x})
\exp\Big(\frac{\langle \boldsymbol{x}, \boldsymbol{s}\star \boldsymbol{x}
 \rangle}{2\sigma_v^2}\Big)
\label{eq:prior.on.x}
\end{align}
we ensure that the \emph{a posteriori} PDFs for the
equalized and target channels are equal:
\[
P(\boldsymbol{X}=\boldsymbol{x}|\boldsymbol{Z}=\boldsymbol{z}) = 
P_T(\boldsymbol{X}=\boldsymbol{x}|\tilde{\boldsymbol{Z}}=\boldsymbol{z})
\]
with the understanding that the left-hand side is the APP
corresponding to the equalized ISI channel (\ref{eq:channel.model})
with a prior $P(\boldsymbol{x})$ on $\boldsymbol{x}$, while the
right-hand side is the APP corresponding to the \emph{target channel}
(\ref{eq:target.channel}) with input PDF $\tilde{P}(\boldsymbol{x})$.
\begin{rmk}
  We reiterate that the target channel is a hypothetical channel and
  we are free to treat its parameters $\boldsymbol{g}$ and
  $\sigma_v^2$ as well as its input PDF $\tilde{P}(\boldsymbol{x})$ as
  \emph{design parameters}.  We assume neither that $\boldsymbol{f}$
  is the MMSE equalizer designed for the target $\boldsymbol{g}$ nor
  that $\sigma_v^2$ is the variance of equalization error. Although
  this approach is radically different from the traditional approach
  in the literature on channel shortening equalization
  \cite{messerschmitt.1974,DBLP:journals/tit/Al-DhahirC96a,lee.hill.1987},
  it is essential to derive the correct form of \emph{equivalence}
  between the target and equalized channels defined below.
\end{rmk}
\begin{defn}
  The equalized channel is equivalent to the target channel in the
  \emph{a posteriori} sense if they produce the same \emph{a
    posteriori} probability for the input given the output. This form
  of equivalence is called \emph{a posteriori} equivalence (APE).
\end{defn}

Evidently, this definition of equivalence is the most natural one from
the perspective of MAP detection. As a caveat, we point out that
$P_T(\tilde{\boldsymbol{Z}}=\boldsymbol{z})$ and
$P({\boldsymbol{Z}}=\boldsymbol{z})$ need not be equal, i.e., the
equalizer output $\boldsymbol{z}$ would not be a typical output of the
target channel. The above observation may be stated succinctly as
follows:
\begin{thm}
The equalized channel (\ref{eq:equalized.channel}) with the prior
distribution $P(\boldsymbol{x})$ and the target channel
(\ref{eq:target.channel}) with the prior distribution
$\tilde{P}(\boldsymbol{x})$ are \emph{a posteriori} equivalent.
\label{thm:A}
\end{thm}

In general, the MMSE or ZFE equalizers do not guarantee this
form of equivalence even though they attempt to make the equalized
channel response close to the target response.

\begin{cor}
  Suppose that the target and equalizer are chosen to be the monic
  solution (\ref{eq:target.for.monic.general}) and
  (\ref{eq:equalizer.for.monic.general}). Furthermore, let $\sigma_v^2
  =\lambda$ and let
\[
\tilde{P}(\boldsymbol{x}) \propto P(\boldsymbol{x})
\exp\Big(
\frac{\langle \boldsymbol{x}, \boldsymbol{s}\star \boldsymbol{x}
 \rangle}{2\lambda}
\Big)
\]
be the input prior distribution for the target channel
(\ref{eq:target.channel}) where $S(\omega)=\lambda/S_x(\omega)$.
Then, the equalized channel is equivalent to the target channel in the
\emph{a posteriori} sense.
\label{cor:MMSEDFE.is.a.solution}
\end{cor}
\begin{proof}
Observe that the monic target (\ref{eq:target.for.monic.general}) and
equalizer (\ref{eq:equalizer.for.monic.general}) satisfy the
hypotheses in Lemma~\ref{lem:part.result} if we set $\alpha =
\lambda/\sigma_w^2$ and $S(\omega)=\lambda/S_x(\omega)$. Therefore, 
by (\ref{eq:sigma_v^2}), $\sigma_v^2 = \alpha\sigma_w^2 =
\lambda$. The claimed result follows from Theorem~\ref{thm:A}.
\end{proof}

The above result shows that we can use the monic design for optimal
MAP detection provided that we use the prior distribution
$\tilde{P}(\boldsymbol{x})$ for the target channel.  In many cases,
the input is IID with a flat spectrum ($S_x(\omega)=1$) implying that
$\tilde{P}(\boldsymbol{x})=P(\boldsymbol{x})$, i.e., we do not need a
different prior PDF for the target channel.
\begin{rmk}
  If we pretend that the equalizer output $\boldsymbol{z}$ came from
  the output of the target channel with a carefully chosen input prior
  distribution, then all MAP-based detection algorithms designed for
  the target channel work optimally when applied to the equalized
  channel. These algorithms include hard-decision decoding such as the
  Viterbi algorithm, and soft-decision decoding such as soft-output
  Viterbi algorithm (SOVA) and the BCJR algorithm. Soft-decision
  algorithms, unlike the Viterbi algorithm, use an extra parameter,
  viz.~the variance of the additive noise in the channel. When
  applying soft decoding to the target channel, we must use
  $\sigma_v^2$ as this variance parameter. Our calculation above show
  that $\sigma_v^2$ simply equals $\lambda$, the equalization error
  variance (see (\ref{eq:white.error})). This fact is routinely
  assumed in many system designs with no rigorous justification but it
  is fortunately the correct value to use.
\label{rmk:zero}
\end{rmk}

\section{Practical Considerations}
\label{sec:practical.considerations}

We now consider some practical implications of our main result in
Section \ref{sec:sequence.detection.equalized}.  Henceforth, we assume
that $P(\boldsymbol{x})$ is a uniform distribution over the set of
allowed code sequences. In this case, the MAP sequence estimate
(\ref{eq:MAP.optimal.rule}) coincides with the ML estimate
(\ref{eq:ML.optimal.rule}).

\begin{thm}
Suppose that all the input sequences in the message codebook have
equal energy and that the equalizer $\boldsymbol{f}$ and target
$\boldsymbol{g}$ are chosen such that
\begin{align}
G^*(\omega)F(\omega) &= \alpha H^*(\omega) \label{eq:G^*F}\\
|G(\omega)|^2 &= \alpha(|H(\omega)|^2+\beta) \label{eq:|G|^2}
\end{align}
for some $\alpha>0$ and $\beta\in\Real$ that produces a valid
$G(\omega)$, then we can set $\tilde{P}(\boldsymbol{x}) =
P(\boldsymbol{x})$. Furthermore, 
if $P(\boldsymbol{x})$ is uniform, the optimal estimate of the input is
\begin{align}
\hat{\boldsymbol{x}} = 
\arg\min_{\boldsymbol{x}} D(\boldsymbol{y}, 
\boldsymbol{x}) = \arg\min_{\boldsymbol{x}} \tilde{D}(\boldsymbol{z}, 
\boldsymbol{x}).
\label{eq:general.equivalence}
\end{align}
\label{thm:equivalence}
\end{thm}
\begin{proof}
In the time domain, the hypotheses imply that
$\boldsymbol{s} = \ddot{\boldsymbol{g}} \star \boldsymbol{g} - 
\alpha \ddot{\boldsymbol{h}} 
\star \boldsymbol{h} =\alpha\beta\boldsymbol{\delta}$ and 
$\ddot{\boldsymbol{g}}\star \boldsymbol{f} =\alpha
\ddot{\boldsymbol{h}}$.  Therefore,
 $\tilde{P}(\boldsymbol{x})={P}(\boldsymbol{x})$. The proof now
readily follows by applying Theorem~\ref{thm:A}.
\end{proof}

Theorem~\ref{thm:equivalence} is applicable, for example, if the input
symbols are elements of the $Q$-phase PSK constellation, i.e.,
$x_n\in\set{C} =\{\sqrt{2}e^{j2\pi q/Q}: q=0,\dots, Q-1\}$ in the
complex case or the BPSK constellation $\set{C} = \{-1, +1\}$ in the
real case, since all message sequences have equal energy.

Clearly, for this special family of equalizer and target filters there
is no performance loss in sequence detection if we minimize the
surrogate cost function $\tilde{D}(\boldsymbol{z}, \boldsymbol{x})$
instead of the original cost $D(\boldsymbol{y},\boldsymbol{x})$.
The practical implications of this result are that in the IIR limit we
can achieve optimal sequence detection using any solution from the
family (see also \cite{raman.icassp06}). In general, these targets are
as long as the channel itself. However, we require a short FIR target
for a Viterbi-based implementation. We address this problem in Section
\ref{sec:FIR.design} where we show how to design good FIR targets to
minimize the detection error rates.

Note that the parameter $\alpha$ is merely a scaling factor (the
target and equalizer scale as $\sqrt{\alpha}$) but $\beta$ affects the
shape of the filters. Thus, we have a degree of freedom in design
represented by $\beta$. We also have the freedom to choose the phase
response of $G(\omega)$.  However, the most logical choice would be to
choose $G(\omega)$ as the causal minimum-phase spectral factor of
(\ref{eq:|G|^2}).  We now consider several interesting cases in the
family of optimal solutions:
\begin{enumerate}
\item The case $\alpha=1$ and $\beta=0$ produces
\[
|G(\omega)|^2=|H(\omega)|^2
\]
and 
\[
F(\omega) = \frac{H^*(\omega)}{G^*(\omega)} = \frac{G(\omega)}{H(\omega)}
\] 
which is an all-pass zero-forcing equalizer filter which keeps the noise
white.
\item Setting $\alpha=\lambda/\sigma_w^2$ and $\beta= \sigma_w^2$
  yields the monic solution (see
  (\ref{eq:MMSE.equalizer.for.monic.IID}) and
  (\ref{eq:target.for.monic.IID})) for $S_x(\omega)=1$, proving its
  conjectured optimality in the asymptotic (IIR) case. When
  $\beta\neq\sigma_w^2$, the solution corresponds to an monic design
  for a different noise level. However, this mismatch causes no
  performance loss in sequence detection. Curiously, some negative
  values $\beta \in (-\inf_{\omega} |H(\omega)|^2, 0)$ also yield
  optimal solutions even though they do not represent the variance of
  any meaningful noise.
\end{enumerate}

\begin{rmk}
  The above argument shows that the monic design is an optimal choice
  if the input spectrum is white. However, suppose that channel input
  spectrum is colored, perhaps by the use of spectral shaping codes.
  Then, the monic design (\ref{eq:equalizer.for.monic.general}) has
  the required form in Theorem \ref{thm:equivalence}. However, the
  target (\ref{eq:target.for.monic.general}) does not because it
  depends on $S_x(\omega)$. Hence, the monic design may be suboptimal
  for colored inputs. In fact, for optimality we must perform the
  monic design for the target and equalizer with an IID input
  regardless of whether the actual input is white or colored. This is
  particularly true at low SNRs where the $\sigma_w$ is large and the
  second term in (\ref{eq:target.for.monic.general}) dominates. At
  high SNR values, the effect of the input spectral color on training
  diminishes.
\label{rmk:one}
\end{rmk}

\subsection{Matched Filter Equalization}
We briefly examine the special case of the solutions in
Theorem~\ref{thm:equivalence} when we let $\beta \to \infty$. This
corresponds to the monic solution for a very low SNR, i.e.,
$\sigma_w^2 \to \infty$. For convenience, we let $\alpha=\beta$
without loss of generality. Then, (\ref{eq:G^*F}) and (\ref{eq:|G|^2})
imply that
\begin{align}
|G(\omega)|^2 = \beta^2 (1+|H(\omega)|^2\beta^{-1})
\label{eq:square}
\end{align}
and 
\[
F(\omega) = \beta H(\omega)^*/G^*(\omega).
\]
For $\beta\gg 1$, we use (\ref{eq:square}) to express $G(\omega)$ as
\[
G(\omega) = \beta + A(\omega)+ O(\beta^{-1})
\]
where $A(\omega)$ must be causal if $\boldsymbol{G}(\omega)$ is
minimum-phase. Thus, as $\beta\to\infty$ we have $F(\omega)$
approaches the \emph{matched filter} $H^*(\omega)$. Now, observe that
\[
|G(\omega)|^2 = \beta^2\Big[1+ (A(\omega)+A^*(\omega))\beta^{-1} + 
O(\beta^{-2})\Big]
\]
Comparing this with (\ref{eq:square}), we obtain
\[
A(\omega)+A^*(\omega) = |H(\omega)|^2 + O(\beta^{-1}).
\]
Therefore, in the time-domain
\[
a_n = \begin{cases} r^{h}_n& \text{if $n>0$}\\ r^{h}_0/2& \text{if $n=0$}
\\ 0, &\text{if $n<0$}
\end{cases}
\]
where $\boldsymbol{r}^h =
\ddot{\boldsymbol{h}}\star \boldsymbol{h}$ is the auto-correlation
function of $\boldsymbol{h}$. 
Using $\boldsymbol{g} = \beta\boldsymbol{\delta} + \boldsymbol{a} 
+ O(\beta^{-1})$ 
it is readily verified that
\begin{align*}
\tilde{D}(\boldsymbol{z}, \boldsymbol{x}) &= \|\boldsymbol{z} - 
\boldsymbol{g}\star\boldsymbol{x} \|^2 \\
&=\|\boldsymbol{z}\|^2 -2\Re \langle 
\boldsymbol{g}\star\boldsymbol{x}, \boldsymbol{z} \rangle + 
\langle \boldsymbol{x}, \ddot{\boldsymbol{g}}\star \boldsymbol{g}\star \boldsymbol{x} \rangle \\
&=\beta^2 \|\boldsymbol{x}\|^2 - 2\beta(\Re \langle \boldsymbol{x},
\boldsymbol{z} \rangle - \langle \boldsymbol{x}, 
\boldsymbol{a}\star \boldsymbol{x}\rangle ) +O(1)
\end{align*}
Since $\|\boldsymbol{x}\|^2$ is constant for all inputs sequences and
$\beta \to \infty$, we deduce that the ML estimation rule becomes
\begin{align}
\arg\min_{\boldsymbol{x}} \tilde{D}(\boldsymbol{z}, \boldsymbol{x}) =
\arg\max_{\boldsymbol{x}} \Re \langle \boldsymbol{x},
\boldsymbol{z} - \boldsymbol{a}\star \boldsymbol{x}\rangle.
\label{eq:first.rule}
\end{align}
We interpret the above calculations as follows. The equalizer is a
matched filter: $\boldsymbol{f} = \ddot{\boldsymbol{h}}$ and the term
$\boldsymbol{z} - \boldsymbol{a}\star \boldsymbol{x}$ represents the
equalizer output with the post-cursor ISI removed using decision
feedback. The estimator simply maximizes the correlation between this
sequence with the input.

It is easy to verify that $\langle \boldsymbol{x} , \boldsymbol{a}
\star \boldsymbol{x} \rangle = \frac{1}{2} \|\boldsymbol{h}\star
\boldsymbol{x} \|^2$. Thus, the matched filter equalization structure
may be derived alternatively directly from Lemma~\ref{lem:part.result}
by letting $\boldsymbol{g} =\boldsymbol{\delta}$ and $\boldsymbol{f} =
\ddot{\boldsymbol{h}}$. This approach gives us the following rule for
ML estimation
\[
\hat{\boldsymbol{x}}= \arg\max_{\boldsymbol{x}} \Re \langle \boldsymbol{x},
\boldsymbol{z} \rangle - \frac{1}{2} \|\boldsymbol{h}\star \boldsymbol{x} \|^2
\]
which is equivalent to (\ref{eq:first.rule}).

\section{Optimal FIR Target Design}
\label{sec:FIR.design}

In the previous sections we showed the existence a family of
equalizers and targets that achieve the optimal sequence detection
performance if we pretend that the equalizer output came from the
target channel.  Unfortunately, the optimal target, being the minimum
phase spectral factor of (\ref{eq:target.for.monic.IID}), has the same
length as the original channel (except in rare cases where it can be
shorter). As such, we have not reduced the detector complexity by
equalization.

In this section, we consider the more practical problem of the design
of FIR targets to achieve the best detection performance.  We consider
only real channels with BPSK input symbols ($\set{C}=\{-1,+1\}$). With
some effort, these result can be generalized to complex channels or
non-binary inputs as well.

Suppose that $\boldsymbol{x}^{\circ}$ is the actual input to the
channel, and $\hat{\boldsymbol{x}}$ is the ML sequence estimate. Then
$\boldsymbol{e} =(\hat{\boldsymbol{x}}-\boldsymbol{x}^{\circ})$ is an
\emph{error sequence}.  We say that two error sequences belong to the
same \emph{equivalence class} if they are related to each other by a
time-shift or phase-rotation (or sign-change).  Of all error
sequences, a \emph{dominant error sequence} is one that which
minimizes $\|\tilde{\boldsymbol{e}}\|^2$ where
$\tilde{\boldsymbol{e}}= \boldsymbol{h}\star \boldsymbol{e}$ is the
noise-free channel response to the input $\boldsymbol{e}$.  We call
$\tilde{\boldsymbol{e}}$ a dominant \emph{output error sequence}.

Clearly, dominant error sequences are not unique because all sequences
in the equivalence class of a dominant error sequences are also
dominant. However, we shall assume that there is a unique dominant
equivalence class whose representative element $\boldsymbol{e}$ has
the canonical form: $e_0\neq 0$ and $e_n=0$ for $n<0$.  Indeed, some
channels could have a multiplicity of dominant events that belong to
the different equivalence classes. In that case our probability of
error estimate would be scaled by the multiplicity factor. 

Let $Q_g(\cdot)$ be the Gaussian $Q$-function
\[
Q_g(x)=\frac{1}{\sqrt{2\pi}}\int_x^\infty e^{-t^2/2}dt.
\]
We now estimate the probability of sequence detection error for any
choice of target and equalizer in terms of the $Q$-function.

\begin{thm}
  At high SNR, the probability of sequence detection error for a real
  BPSK channel is given by $P^{\rm seq}_e \simeq \kappa Q_g(\sqrt{\SNR})$
  for some constant $\kappa$ with $\SNR$ is the \emph{effective
  signal-to-noise ratio} of the system
\begin{align*}
\SNR=\max_{\boldsymbol{v}} 
\frac{|\Re \langle \boldsymbol{e},
\boldsymbol{p}\star \boldsymbol{h} \star 
\boldsymbol{e} \rangle|^2}{{\|(\boldsymbol{q} - 
\ddot{\boldsymbol{p}}\star 
\ddot{\boldsymbol{h}})\star \boldsymbol{e} -\boldsymbol{v}\|^2
+\sigma_w^2\|\boldsymbol{p}\star\boldsymbol{e}\|^2}}
\end{align*}
where $\boldsymbol{p} = \boldsymbol{f}\star \ddot{\boldsymbol{g}}$,
$\boldsymbol{q} =\boldsymbol{g}\star \ddot{\boldsymbol{g}}$, and
$\boldsymbol{v}$ is any sequence with the same temporal support as the
dominant error sequence $\boldsymbol{e}$.
\label{thm:prob.err}
\end{thm}

Theorem~\ref{thm:prob.err} is proved in Appendix \ref{app:A} using
error analysis similar to that of standard Viterbi detection
\cite{book:proakis,forney.1972}.  Note that the bit error rate (BER)
also takes the same form as $P_e^{\rm seq}$ but has a different
constant than $\kappa$. The above result is applicable for FIR and IIR
equalizers and targets. The optimal equalizer
$\boldsymbol{f}$ and target $\boldsymbol{g}$ are chosen to maximize
$\SNR$ subject to relevant constraints.

For practical reasons, we seek FIR targets, since the detector
implementation complexity is exponential in the target length. The
constraint on the equalizer length is less important since the
complexity growth is only linear. For simplicity we assume that the
equalizer is IIR but the target is FIR with length $L$.  In this case,
it is more convenient to maximize $\SNR$ over $\boldsymbol{p}$ and
$\boldsymbol{q}$ because $\boldsymbol{f}$ and $\boldsymbol{g}$ can be
recovered uniquely from $\boldsymbol{p}$ and $\boldsymbol{q}$ by
spectral factorization.  Note that $\boldsymbol{p}$ is IIR but
$\boldsymbol{q}$, being the autocorrelation function of
$\boldsymbol{g}$, is FIR. Furthermore, we have $Q(\omega)\geq 0$.  We
write $\SNR=\max_{\boldsymbol{v}} \SNR(\boldsymbol{p}, \boldsymbol{q},
\boldsymbol{v})$ where
\[ 
\SNR(\boldsymbol{p}, \boldsymbol{q}, \boldsymbol{v})\defas
\frac{|\Re \langle \boldsymbol{e},
\boldsymbol{p}\star \boldsymbol{h} \star 
\boldsymbol{e} \rangle|^2}{{\|(\boldsymbol{q} - 
\ddot{\boldsymbol{p}}\star 
\ddot{\boldsymbol{h}})\star \boldsymbol{e} -\boldsymbol{v}\|^2
+\sigma_w^2\|\boldsymbol{p}\star\boldsymbol{e}\|^2}}
\]
Now observe that
\[
\SNR(\boldsymbol{p}, \boldsymbol{q}, \boldsymbol{v}) =
\SNR(\boldsymbol{p}, \boldsymbol{q}+\beta\boldsymbol{\delta}, \boldsymbol{v}-
\beta\boldsymbol{e})
\]
for any $(\boldsymbol{p}, \boldsymbol{q}, \boldsymbol{v})$ and $\beta
\in \Real$. Moreover, if $\boldsymbol{v}$ has the same temporal
support as $\boldsymbol{e}$, then so does $\boldsymbol{v}'=
\boldsymbol{v}- \beta\boldsymbol{e}$. Since we are minimizing
$\SNR(\boldsymbol{p}, \boldsymbol{q}, \boldsymbol{v})$ over all
$\boldsymbol{v}$, we conclude that the quantity
\begin{align}
\max_{\boldsymbol{p}, \boldsymbol{v}} \SNR(\boldsymbol{p}, 
\boldsymbol{q}, \boldsymbol{v}).
\label{eq:filter.design}
\end{align}
would remain unchanged if we replace $\boldsymbol{q}$ by
$\boldsymbol{q} + \beta
\boldsymbol{\delta}$. This enables us to temporarily replace constraint
$Q(\omega)\geq 0$ by $q_0=0$ for the sake of the maximization.  Having
rid of the constraint on $Q(\omega)$, the maximization is readily
transformed into a quadratic minimization.  As a final step, we add a
sufficiently large $\beta$ to the solution $Q(\omega)$ to make it
satisfy $Q(\omega)\geq 0$.

The analytical solution to (\ref{eq:filter.design}) is presented in
the Appendix \ref{app:B}. We also show there that the noise variance
in the hypothetical target channel noise variance (\ref{eq:sigma_v^2})
is set to $\sigma_v^2=\lambda$, the Lagrange multiplier used in the
optimization.

Clearly, the above maximization admits infinitely many solutions
parameterized by $\beta$. As the target length approaches infinity,
these solutions converge precisely to the family of solutions in
Theorem~\ref{thm:equivalence}. In this limit the equalizer and
target filters of Theorem~\ref{thm:A} maximize the effective
SNR. Furthermore, this maximum value is
\begin{align}
\SNR_{\max}=\frac{\|\boldsymbol{h}\star \boldsymbol{e}\|^2}{\sigma_w^2}.
\label{eq:snr.max}
\end{align}

In practice we are interested in FIR equalizers for ease of
implementation.  We point out that we could still maximize the effective 
SNR, albeit numerically, over all FIR targets and equalizers
with length constraints. If we choose to use FIR equalizers, we would
have the additional task of optimizing the processing delay which is
an important design parameter
\cite{DBLP:journals/tit/Al-DhahirC96a,dhahir.cioffi.1995,dhahir.2000}.

\section{Examples}
\label{sec:examples}

We now illustrate our results of the preceeding sections with an example.
Consider the real ISI channel (\ref{eq:channel.model}) with impulse response
\[
h_n= \begin{cases}
e^{-n/2},& 0 \leq n \leq 8\\
0 & \text{otherwise.}
\end{cases}
\]
with IID binary input symbols ($x_n \in \set{C} = \{ -1, +1\}$) and SNR
defined as ${\|\boldsymbol{h}\|^2}/{\sigma_w^2}$ where $\sigma_w^2$ is
the noise variance.

\begin{figure}
\begin{minipage}[b]{1.0\linewidth}
  \centering
  \centerline{\epsfig{figure=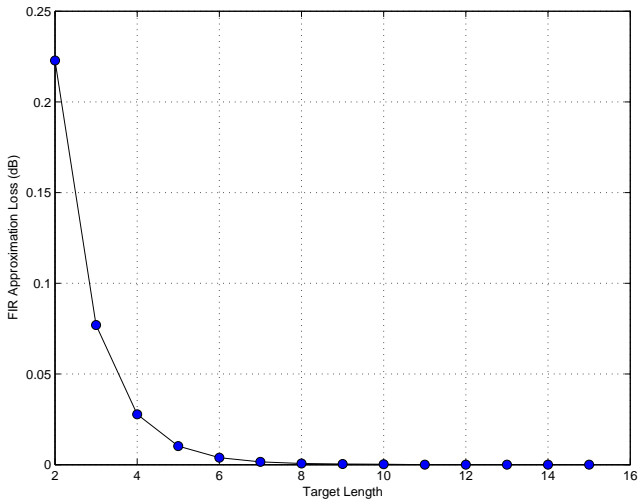,width=\FIGWIDTH}}
  \caption{FIR approximation loss vs.\ target length}
\label{fig:FIR.loss}
\end{minipage}
\end{figure}

We first study the effect of the target length on the effective SNR of
the system.  The optimal equalizers and targets are computed for
target lengths of 2 and longer and the resulting values of $\SNR$ are
calculated. Indeed, in the IIR limit for the target length we obtain
the maximum value $\SNR_{\max}$ given by (\ref{eq:snr.max}).
Fig.~\ref{fig:FIR.loss} shows the \emph{FIR approximation loss},
$(\SNR_{\max} - \SNR)$, for various finite target lengths at an SNR of
10dB. In this example the optimal length-3 target incurs about 0.075dB
penalty in performance and the performance loss for longer targets
diminishes quickly.

Next, we evaluate the BER performance of the reduced complexity
detectors. At each SNR we design the optimal length-3 target and IIR
equalizer truncated to 21-taps (centered at the origin). The equalizer is
sufficiently long since it
captures most of the energy in the equalizer taps.  The dominant error
event for this channel is $\boldsymbol{e}=\{1, -1\}$. We also design
length-21 MMSE equalizers (centered at 0) and length-3 targets
described in Section \ref{sec:linear.equalization} for the 
\emph{monic} target constraint.

\begin{figure}[htb]
\begin{minipage}[b]{1.0\linewidth}
  \centering
  \centerline{\epsfig{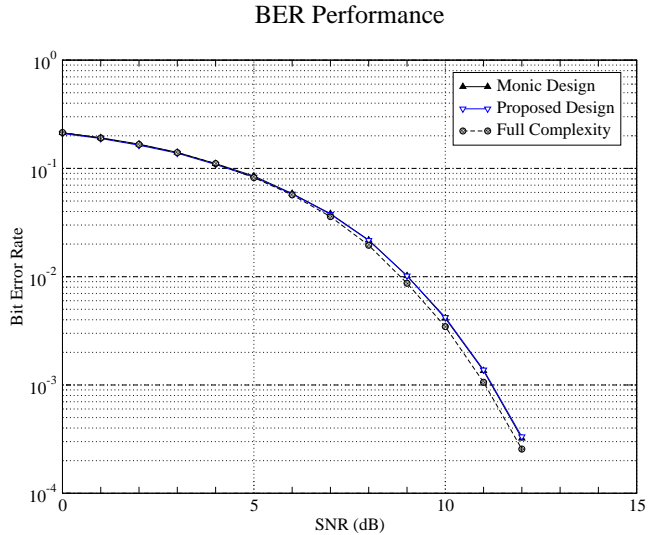}}
  \caption{Comparison of BER performance of two 
    designs for binary signaling}
  \label{fig:binary.BER}
\end{minipage}
\end{figure}

Using computer simulations we compare the two designs in terms of
their BER performance for IID binary inputs. The two systems use the
Viterbi algorithm to perform the sequence detection. The results are
shown in Fig.~\ref{fig:binary.BER} along with the BER of the full
complexity Viterbi detector (with $2^8$-states) that uses no channel
shortening equalization.  It is clear that both the reduced complexity
detectors performanc identically with a small penalty
relative to the full complexity detector. The optimality of the monic 
design is predicted by Theorem~\ref{thm:equivalence} for the case 
of IIR filters.  Indeed, we observe numerically that the monic design
is nearly optimal for FIR filters as well.

Next, we consider the same ISI chanel with an IID ternary input ($x_n \in \set{C}
= \{ -\sqrt{3/2}, 0, +\sqrt{3/2}\}$) which has unit average symbol energy.  
This input symbols themselves have unequal energy.
Recall the results for the IIR case in Section \ref{sec:sequence.detection.equalized} that
the optimal sequence detector for the equalized channel needs to
pretend that it sees the output of the \emph{target channel} with the
input prior distribution is given by (\ref{eq:prior.on.x}):
\begin{align*}
\tilde{P}(\boldsymbol{x}) &\propto 
\exp\Big(\frac{\langle \boldsymbol{x}, \boldsymbol{s}\star \boldsymbol{x}
 \rangle}{2\sigma_v^2}\Big)
\end{align*}
where $\boldsymbol{s} =\ddot{\boldsymbol{g}} \star \boldsymbol{g} -
\alpha \ddot{\boldsymbol{h}}\star\boldsymbol{h}$. Thus, 
the optimal detector needs to minimize the cost function
\[
\min_{\boldsymbol{x}} \big( 
\|\boldsymbol{z}-\boldsymbol{g}\star\boldsymbol{x}\|^2 - 
\langle \boldsymbol{x}, \boldsymbol{s}\star \boldsymbol{x}
 \rangle \big)
\]
where the second term is correction term that originates from the
input prior distribution $\tilde{P}(\boldsymbol{x})$.  For the choice
of equalizer and target in Theorem~\ref{thm:equivalence}, we have
\[
\boldsymbol{s} \defas
\ddot{\boldsymbol{g}} \star \boldsymbol{g} -
\alpha \ddot{\boldsymbol{h}} 
\star \boldsymbol{h} =\alpha\beta\boldsymbol{\delta}.
\]
Therefore, $\langle \boldsymbol{x},\boldsymbol{s}\star \boldsymbol{x}
\rangle = \alpha\beta\|\boldsymbol{x}\|^2$, which depends on the
energy of the sequence. The correction term is an issue only for
signal constellations unequal symbol energies. For the monic target
and MMSE equalizer design, we have $\alpha\beta$ equals the variance
of the equalization error, $\lambda$. Thus, the cost function reduces
to
\[
\min_{\boldsymbol{x}} \big( 
\|\boldsymbol{z}-\boldsymbol{g}\star\boldsymbol{x}\|^2 - 
\lambda \|\boldsymbol{x}\|^2 \big).
\]
We directly adapt this expression to the FIR case as well by
subtracting $\lambda |x_n|^2$ from the trellis branch metric at time
$n$. In fact, the detector would be suboptimal without the correction
term, as we confirm below.

\begin{figure}[htb]
\begin{minipage}[b]{1.0\linewidth}
  \centering 
  \centerline{\epsfig{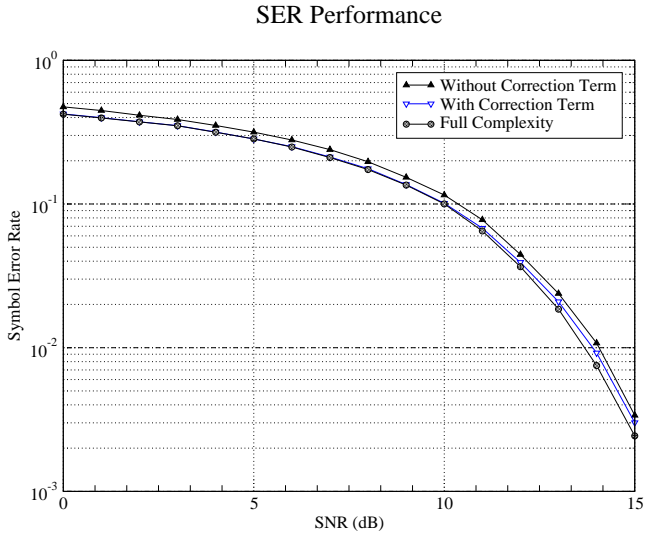}}
  \caption{Comparison of SER performance for ternary input signaling}
  \label{fig:ternary.SER}
\end{minipage}
\end{figure}

We design a length-3 monic GPR target and a length-21 MMSE equalizer
for this channel and calculate the symbol error rates (SER)
numerically using the Viterbi algorithm. Fig.~\ref{fig:ternary.SER}
shows the SER obtained with and without the correction term in the trellis 
branch metric. The figure also shows the SER for the full complexity Viterbi 
detector (with $3^8$ states) that uses no channel shortening equalization.  
There is a small but noticeable gain in detection performance with the 
correction term. It must be noted that this modification does not require 
much more detector complexity. As $\lambda$ becomes smaller (at higher
SNRs) the correction term to becomes smaller also and indeed, the
performances gain due to the correction term diminishes at high SNRs.

\section{summary}

Although a large body of literature exists for the design of optimal
FIR targets and equalizers, the implicit assumption in virtually all
existing work on this subject is that MMSE equalization is optimal.  
The purpose of this work was to question that assumption.
The main contribution of this work is a new perspective for the
problem of channel shortening equalization in terms of the underlying
\emph{a posteriori} probabilities unlike the traditional
approach of using the MSME equalization error as the criterion.  We
introduced the idea of \emph{a posteriori} equivalence (APE) between
the equalized and target channels. Under this form of
equivalence, any MAP-based decoding algorithm designed for the target
channel would also work \emph{optimally} when applied to the equalized 
channel. In other words, as far as MAP decoding is concerned we can
pretend that the equalized channel \emph{is} the target channel.

In our analysis of the problem we treat $\boldsymbol{f}$,
$\boldsymbol{g}$, $\sigma_v^2$ (noise variance in the target channel)
and in some cases even the input PDF $\tilde{P}(\boldsymbol{x})$ for
the hypothetical target channel as design parameters.  The equivalence
is expressed as a set of algebraic conditions on the design
parameters.  The APE conditions admit an infinite family solutions or
designs for the equalizer and target.  In the special case that the
input is IID and all the code sequences have equal energy, we showed
that the ``monic solution,'' i.e., the MMSE equalizer designed for a
monic constrained target, is shown to belong to this optimal family of
designs. We also observed that the monic solution must be designed for
spectrally white inputs even if the actual input is colored.  The
family of designs produces IIR filters in general, making their
practical use somewhat limited, where as for a low complexity
implementation of optimal sequence detection (using Viterbi or
BCJR-like algorithms) we require short FIR targets.

We also derived an expression for the probability of sequence
detection error assuming IID inputs for arbitrary FIR or IIR targets
and equalizers. Using this as a performance measure, we propose a
design algorithm to find the optimal IIR equalizer and FIR target.
Indeed, in the IIR limit for the target these solution coincide with
the previously derived optimal IIR family of designs that satisfy
APE.

These results are applied to an example ISI channel. Numerically, we
observe that for IID inputs, we obtain nearly optimal performance
using the monic design.  for input signal constellations with unequal
symbol energies we also
need to treat the input PDF $\tilde{P}(\boldsymbol{x})$ for the target
channel as a design parameter. The optimal detector is designed for
the target channel with the prior $\tilde{P}(\boldsymbol{x})$
incorporated into the Viterbi branch metric as a correction term, which 
would normally have been ignored if we simply use the monic design.
This is illustrated for the IID ternary signaling example
(Fig.~\ref{fig:ternary.SER}) where we see a small but noticeable gain
by using the correction term.

\appendices
\section{Proof of Theorem~\ref{thm:prob.err}}
\label{app:A}

Suppose that $\boldsymbol{x}^\circ\in\set{X}$ is the transmitted
sequence, where $\set{X}$ is the set of sequences that are equally
likely to be transmitted. The channel and equalizer outputs are
$\boldsymbol{y}=\boldsymbol{h}\star \boldsymbol{x} + \boldsymbol{w}$
and $\boldsymbol{z}=\boldsymbol{f}\star\boldsymbol{y}$ respectively.
All sequences in the codebook have equal energy because the input
symbols are IID and binary. Thus, the target channel input is also
treated as being IID: $\tilde{P}(\boldsymbol{x})=P(\boldsymbol{x})$.

The Viterbi detector for the equalized channel computes the sequence
$\boldsymbol{x}$ that minimizes $\tilde{D}(\boldsymbol{z},
\boldsymbol{x})$. Thus, the probability of sequence detection
error is
\begin{align}
P_e^{\rm seq} 
&= P\big\{\tilde{D}(\boldsymbol{z}, \boldsymbol{x}) <
\tilde{D}(\boldsymbol{z}, \boldsymbol{x}^\circ)\text{ for some
$\boldsymbol{x}\neq \boldsymbol{x}^\circ$} \big\} \nonumber \\
&\leq \frac{1}{|\set{X}|}
\sum_{\boldsymbol{x}^\circ \in \set{X}} \sum_{\boldsymbol{x} \in \set{X} \backslash 
\boldsymbol{x}^\circ}
 P\big\{\tilde{D}(\boldsymbol{z}, \boldsymbol{x}) <
\tilde{D}(\boldsymbol{z}, \boldsymbol{x}^\circ)\big\}
\label{eq:union.bound}
\end{align}
where the second step follows from the union bound.
Using the property that 
\begin{align}
\|\boldsymbol{a}\|^2 -\|\boldsymbol{b}\|^2 =
\Re \langle \boldsymbol{a}- \boldsymbol{b},
\boldsymbol{a}+ \boldsymbol{b} \rangle
\label{eq:property}
\end{align}
for any $\boldsymbol{a}$ and $\boldsymbol{b}$, where $\Re$ denotes the
real part, we obtain
\begin{align}
\tilde{D}(\boldsymbol{z}, \boldsymbol{x}) - 
\tilde{D}(\boldsymbol{z}, \boldsymbol{x}^\circ) 
&= \|\boldsymbol{z} - \boldsymbol{g}\star
\boldsymbol{x} \|^2 - \|\boldsymbol{z} - 
\boldsymbol{g}\star\boldsymbol{x}^\circ \|^2 \nonumber \\
&=-4\Re \langle \boldsymbol{g}\star \boldsymbol{x}^-, 
\boldsymbol{z} - \boldsymbol{g}\star
\boldsymbol{x}^+ \rangle 
\label{eq:difference}
\end{align}
where
\[
\boldsymbol{x}^\pm \defas \frac{\boldsymbol{x} \pm \boldsymbol{x}^\circ}{2}.
\]

Applying (\ref{eq:adjoint}) to (\ref{eq:difference}) 
and writing $\boldsymbol{z}=\boldsymbol{f}\star \boldsymbol{y}$ 
where $\boldsymbol{y} = \boldsymbol{h}\star (\boldsymbol{x}^+ -
\boldsymbol{x}^-) +\boldsymbol{w}$ 
we obtain
\begin{align*}
\tilde{D}(\boldsymbol{z}, \boldsymbol{x}) - 
\tilde{D}(\boldsymbol{z}, \boldsymbol{x}^\circ) 
&=4\Re \langle \boldsymbol{x}^-,
\ddot{\boldsymbol{g}}\star \boldsymbol{f}\star \boldsymbol{h} \star 
\boldsymbol{x}^- \rangle \\
&\quad  + 4\Re \langle \boldsymbol{x}^-,
\ddot{\boldsymbol{g}}\star (\boldsymbol{g}- \boldsymbol{f}\star \boldsymbol{h})
 \star \boldsymbol{x}^+ \rangle \\
&\quad - 4\Re \langle \boldsymbol{x}^-,
\ddot{\boldsymbol{g}}\star \boldsymbol{f}\star \boldsymbol{w} \rangle \\
&\equiv 4(\phi(\boldsymbol{x}^-) + \Delta(\boldsymbol{x}^-,
\boldsymbol{x}^+) - \psi(\boldsymbol{x}^-))
\end{align*}
where
\begin{align}
\phi(\boldsymbol{x}^-) &\defas \Re \langle \boldsymbol{x}^-,
\boldsymbol{p}\star \boldsymbol{h} \star 
\boldsymbol{x}^- \rangle \label{eq:phi} \\
\Delta(\boldsymbol{x}^-,\boldsymbol{x}^+) &\defas \Re \langle \boldsymbol{x}^-,
(\boldsymbol{q}- \boldsymbol{p}\star \boldsymbol{h})
 \star \boldsymbol{x}^+ \rangle \label{eq:Delta} \\
\psi(\boldsymbol{x}^-) &\defas \Re \langle \boldsymbol{x}^-,
\boldsymbol{p}\star \boldsymbol{w} \rangle
\label{eq:psi}\\
\boldsymbol{p} &\defas \ddot{\boldsymbol{g}} \star \boldsymbol{f}\\
\boldsymbol{q} &\defas \ddot{\boldsymbol{g}} \star \boldsymbol{g}.
\end{align}

Note that $\psi(\boldsymbol{x}^-)\sim N(0,
\sigma_w^2\|{\boldsymbol{p}}\star
\boldsymbol{x}^-\|^2)$ is normally distributed. Therefore,
\begin{align}
\Pi(\boldsymbol{x}^-, \boldsymbol{x}^+) &\defas 
P\big\{\tilde{D}(\boldsymbol{z}, \boldsymbol{x}) <
\tilde{D}(\boldsymbol{z}, \boldsymbol{x}^\circ)\big\} \nonumber \\
&=P\big\{ \psi(\boldsymbol{x}^-) - \Delta(\boldsymbol{x}^-,
\boldsymbol{x}^+) > \phi(\boldsymbol{x}^-) \big\}.
\label{eq:Pi.defn}
\end{align}

Thus, (\ref{eq:union.bound}) can be rewritten as
\begin{align*}
P_e^{\rm seq}
&\leq 
\frac{1}{|\set{X}|} \sum_{\boldsymbol{x}^-\neq 0} 
\sum_{\boldsymbol{x}^+ \in \set{X}^+(\boldsymbol{x}^-)}
\Pi(\boldsymbol{x}^-, \boldsymbol{x}^+)
\end{align*}
where $\set{X}^+(\boldsymbol{x}^-)$ is the set of sequences
$\boldsymbol{x}^+$ such that $\boldsymbol{x}^+ + \boldsymbol{x}^-$ and
$\boldsymbol{x}^+ - \boldsymbol{x}^-$ are valid sequences
in $\set{X}$.  Note that $\boldsymbol{x}^+$ is uniformly distributed in
$\set{X}^+(\boldsymbol{x}^-)$ when conditioned on $\boldsymbol{x}^-$.
In the high SNR regime, it is a good approximation to assume that
dominant error sequences are the only source of detection errors. This
allows us to fix $\boldsymbol{x}^-=\boldsymbol{e}$ for any error
sequence $\boldsymbol{e}\in \set{E}$ in the equivalence class
$\set{E}$ of dominant error sequences.  This yields
\begin{align*}
P_e^{\rm seq}
&\leq \frac{|\set{E}|}{|\set{X}|} 
\sum_{\boldsymbol{x}^+} 
\Pi(\boldsymbol{e}, \boldsymbol{x}^+)\\
&= \frac{|\set{E}||\set{X}^+(\boldsymbol{e})|}{|\set{X}|} 
\rE \Pi(\boldsymbol{e}, \boldsymbol{x}^+)
\end{align*}
with the expectation taken over $\boldsymbol{x}^+$ given
that $\boldsymbol{x}^-=\boldsymbol{e}$. For analytical tractability, we
assume that $\Delta(\boldsymbol{e},\boldsymbol{x}^+)$ is approximately
normally distributed. Thus, (\ref{eq:Pi.defn}) yields
\begin{align}
P_e^{\rm seq} \simeq \kappa Q\Big(
\frac{\phi(\boldsymbol{e})}{\sigma(\boldsymbol{e})}\Big)
\label{eq:Pseq}
\end{align}
where
\begin{align}
\sigma^2(\boldsymbol{e}) &= {\var (\Delta(\boldsymbol{e},
\boldsymbol{x}^+))+
\sigma_w^2\|\boldsymbol{p}\star\boldsymbol{x}^-\|^2}
\label{eq:sigma^2}\\
\kappa &= |\set{E}|\frac{|\set{X}^+(\boldsymbol{e})|}{|\set{X}|} .
\label{eq:kappa}
\end{align}
The constant $\kappa$ is evidently the product of the number of
allowable dominant error sequences $|\set{E}|$ and the probability,
$|\set{X}^+(\boldsymbol{e})|/|\set{X}|$, that $\boldsymbol{x}^\circ$
will allow that error sequence.  The bit error rate (BER) is
approximated by
\begin{align}
P_e^{\rm bit} = \frac{w_H(\boldsymbol{e})}{M}P_e^{\rm seq}
\label{eq:BER.approximation}
\end{align}
where $M$ is the length of the input codewords.  The above
calculations are similar to probability of error analysis for
classical Viterbi detection \cite{book:proakis}.

The only remaining step is to estimate the variance of
$\Delta(\boldsymbol{e},\boldsymbol{x}^+)$. 
First note that 
\begin{align*}
\Delta(\boldsymbol{e},\boldsymbol{x}^+) &= \Re \langle \boldsymbol{e},
(\boldsymbol{q}- \boldsymbol{p}\star \boldsymbol{h})
 \star \boldsymbol{x}^+ \rangle \\ &= \Re \langle \boldsymbol{a},
\boldsymbol{x}^+ \rangle
\end{align*}
where $\boldsymbol{a}= (\boldsymbol{q} - \ddot{\boldsymbol{p}}\star
\ddot{\boldsymbol{h}})\star \boldsymbol{e}$.  
Now, $\Delta(\boldsymbol{e},\boldsymbol{x}^+)$ is zero-mean because
$\boldsymbol{x}^+$ is zero-mean.  Hence, the conditional variance of
$\Delta(\boldsymbol{e},\boldsymbol{x}^+)$ is

\begin{align*}
\var (\Delta(\boldsymbol{e},\boldsymbol{x}^+)) 
&=\frac{1}{|\set{X}^+(\boldsymbol{e})|} 
\sum_{\boldsymbol{x}^+ \in \set{X}^+(\boldsymbol{e})}
(\Delta(\boldsymbol{e},\boldsymbol{x}^+))^2\\
&=\frac{1}{|\set{X}^+(\boldsymbol{e})|} 
\sum_{\boldsymbol{x}^+ \in \set{X}^+(\boldsymbol{e})}
|\Re \langle\boldsymbol{a},\boldsymbol{x}^+ \rangle|^2.
\end{align*}

Since the input is binary with symbols being $\pm 1$,
$\set{X}^+(\boldsymbol{x}^-)$ contains all sequences
$\boldsymbol{x}^+$ that satisfy
\[
e_n\neq 0 \implies x^+_n =0.
\]
It is an easy exercise to check that
\begin{align*}
\var(\Delta(\boldsymbol{e},\boldsymbol{x}^+)) 
&=\sum_{\{n:e_n=0\}}|a_n|^2
\end{align*}
which may also be written as
\begin{align}
\var(\Delta(\boldsymbol{e},\boldsymbol{x}^+)) 
&=\min_{\boldsymbol{v}} \|\boldsymbol{a}-\boldsymbol{v}\|^2 \nonumber \\
&= \min_{\boldsymbol{v}} \|(\boldsymbol{q} - 
\ddot{\boldsymbol{p}}\star 
\ddot{\boldsymbol{h}})\star \boldsymbol{e} -\boldsymbol{v}\|^2
\label{eq:var.Delta}
\end{align}
where $\boldsymbol{v}$ is a vector whose temporal support is the same
as that of $\boldsymbol{e}$. Combining (\ref{eq:phi}),
(\ref{eq:Pseq}), (\ref{eq:sigma^2}), and (\ref{eq:var.Delta}), we
obtain $P_e^{\rm seq} \simeq \kappa Q_g(\sqrt{\SNR})$ where
\begin{align*}
\SNR &\simeq \frac{|\Re \langle \boldsymbol{e},
\boldsymbol{p}\star \boldsymbol{h} \star 
\boldsymbol{e} \rangle|^2}{{
\min_{\boldsymbol{v}} \|(\boldsymbol{q} - 
\ddot{\boldsymbol{p}}\star 
\ddot{\boldsymbol{h}})\star \boldsymbol{e} -\boldsymbol{v}\|^2
+\sigma_w^2\|\boldsymbol{p}\star\boldsymbol{e}\|^2}} \\
&=\max_{\boldsymbol{v}} 
\frac{|\Re \langle \boldsymbol{e},
\boldsymbol{p}\star \boldsymbol{h} \star 
\boldsymbol{e} \rangle|^2}{{\|(\boldsymbol{q} - 
\ddot{\boldsymbol{p}}\star 
\ddot{\boldsymbol{h}})\star \boldsymbol{e} -\boldsymbol{v}\|^2
+\sigma_w^2\|\boldsymbol{p}\star\boldsymbol{e}\|^2}}
\end{align*}
is the \emph{effective SNR} of the system. \hfill \QED

\section{Analytical Solution to (\ref{eq:filter.design})}
\label{app:B}

The maximization (\ref{eq:filter.design}) may be rewritten as
\[ 
\min_{\boldsymbol{p}, \boldsymbol{q}, \boldsymbol{v}}
{{\|(\boldsymbol{q} - 
\ddot{\boldsymbol{p}}\star 
\ddot{\boldsymbol{h}})\star \boldsymbol{e} -\boldsymbol{v}\|^2
+\sigma_w^2\|\boldsymbol{p}\star\boldsymbol{e}\|^2}}
\]
subject to $q_0=0$ and 
\begin{align}
\Re\langle \boldsymbol{e},
\boldsymbol{p}\star \boldsymbol{h} \star 
\boldsymbol{e} \rangle = 1
\label{eq:additional.constraint}
\end{align}
thereby removing the scaling invariance of the solutions. Define $\set{S}
=\{l: e_l\neq 0\} =\{s_1,\dots, s_J\}$. Then
\begin{align*}
V(\omega) &= \sum_{l\in\set{S}} v_l e^{-jl\omega} \\
Q(\omega) &= 2\sum_{l=1}^L q_l \cos(l\omega).
\end{align*}
where $v_l$, $l\in\set{S}$ and $q_l: l=1,\dots, L$ are the FIR
parameters. Therefore,
\[
Q(\omega)E(\omega) - V(\omega) = \boldsymbol{B}(\omega) \boldsymbol{x}
\]
where
\begin{align*}
\boldsymbol{B}(\omega) &= 
\begin{pmatrix} \boldsymbol{B}_1(\omega) & 
\boldsymbol{B}_2(\omega) \end{pmatrix} \\
\boldsymbol{B}_1(\omega) &= 2E(\omega)
\begin{pmatrix} \cos(\omega) &
\cos(2\omega) & \cdots & \cos(L\omega) 
\end{pmatrix} \\
\boldsymbol{B}_2(\omega) &= -\begin{pmatrix} e^{-js_1\omega}, & \cdots, & 
e^{-js_J\omega} \end{pmatrix}\\
\boldsymbol{x} &= \begin{pmatrix} q_1, & \cdots, & q_L, & v_{s_1}, & 
\cdots, & v_{s_J} \end{pmatrix}^T
\end{align*}
Finally, let
\begin{align}
R(\omega) \defas P(\omega)H(\omega) E(\omega).
\label{eq:transformation}
\end{align}
In terms of the above quantities, we can rewrite the optimization as
\begin{align*}
\min \frac{1}{2\pi}\Big[
\int |\boldsymbol{B}(\omega) \boldsymbol{x} - R(\omega)|^2 d\omega + 
{\sigma_w^2}
\int | R(\omega)/H(\omega)|^2 d\omega \Big]
\end{align*}
subject to
\begin{align}
\frac{1}{2\pi} \Re \int R^*(\omega) E(\omega) d\omega = 1.
\label{eq:constraint}
\end{align}
All integrals are taken over $[-\pi,\pi]$.  The cost function reduces to
\begin{align*}
\min \Big[
\frac{1}{2\pi} \Re \int \Big(A(\omega) |R(\omega)|^2  - 
2 R(\omega)^* \boldsymbol{B}(\omega)\boldsymbol{x}\Big) d\omega 
+
\boldsymbol{x}^*
\boldsymbol{C}\boldsymbol{x}\Big]
\end{align*}
where $A(\omega) = 1 + \sigma_w^2/|H(\omega)|^2$ and 
\[
\boldsymbol{C} = \frac{1}{2\pi}\int \boldsymbol{B}^*(\omega) 
\boldsymbol{B}(\omega) d\omega.
\]
Using variational calculus we obtain
\begin{align*}
A(\omega)R(\omega) - \boldsymbol{B}(\omega)\boldsymbol{x} &= \lambda E(\omega)
\\ - \frac{1}{2\pi} \int \boldsymbol{B}^*(\omega)R(\omega) d\omega +
\boldsymbol{C} \boldsymbol{x} &= \boldsymbol{0}
\end{align*}
where $\lambda$ is a Lagrange multiplier.  Solving the above
simultaneous equations yields
\begin{align*}
R(\omega) &= \frac{\boldsymbol{B}(\omega)\boldsymbol{x} +
\lambda E(\omega)}{A(\omega)} \\
\boldsymbol{x} &=  \lambda
( \boldsymbol{C}-\boldsymbol{D})^{-1} \int 
\frac{\boldsymbol{B}^*(\omega) E(\omega)}{2\pi A(\omega)}d\omega
\end{align*}
where
\[
\boldsymbol{D}=\frac{1}{2\pi} \int \frac{\boldsymbol{B}^*(\omega) 
\boldsymbol{B}(\omega)}{A(\omega)}d\omega.
\]
Finally $P(\omega)$ can be solved from (\ref{eq:transformation}). Note
that $\lambda$ is uniquely determined by the constraint
(\ref{eq:constraint}). However, we could choose an arbitrary value for
$\lambda$ (such as $\lambda=1$) since it merely scales the solution
without altering the value of $\SNR$.

In the long target (IIR) limit, it is easy to see that the solutions
converge to the following limits:
\begin{align*}
\boldsymbol{f}\star \ddot{\boldsymbol{g}} &= 
\boldsymbol{p} \to \frac{\lambda\ddot{\boldsymbol{h}}}{\sigma_w^2},\\
\ddot{\boldsymbol{g}}\star \boldsymbol{g} &= \boldsymbol{q} \to 
\frac{\lambda(\boldsymbol{h}\star \ddot{\boldsymbol{h}}+\beta)}{\sigma_w^2}
\end{align*}
for some $\beta$, which have the required form in
Theorem~\ref{thm:equivalence} for $\alpha=\lambda/\sigma_w^2$.
Therefore, (\ref{eq:sigma_v^2}) suggests that we must set
\[
\sigma_v^2 = \alpha \sigma_w^2 = \lambda.
\]
In the FIR case, however, the problem of choosing the correct value of
$\sigma_v^2$ is somewhat ambiguous because FIR solutions do not
satisfy the hypotheses in Theorem~\ref{thm:equivalence}. We nominally
set $\sigma_v^2 = \lambda$ in the FIR case as well. This is a good
first approximation and fine-tuning this parameter may produce better
results.


\bibliographystyle{IEEEbib}

\end{document}